\documentclass[useAMS,usegraphicx,usenatbib]{mn2e}
\usepackage[a4paper,centering, totalwidth=520pt, totalheight=700pt]{geometry}
\usepackage{graphicx}
\usepackage{aas_macros}     
\usepackage{amsmath}
\usepackage{amssymb}
\usepackage{fixltx2e}[1999/12/01]

\newcommand{\Mass}{ \ensuremath{ h^{-1} M_{\odot}} }
\newcommand{\Mpc}{ \ensuremath{h^{-1} {\rm Mpc}} }
\newcommand{\Gpc}{ \ensuremath{h^{-1} {\rm Gpc}} }
\newcommand{\hMpc}{ \ensuremath{ h {\rm Mpc^{-1}}} }
\newcommand{\kpc}{ \ensuremath{ h^{-1} {\rm kpc}} }

\title[How closely do baryons follow DM on large scales?]{How closely do baryons follow dark matter on large scales?}

\begin{document}
\setlength{\topmargin}{-1.cm}

\author[Angulo et al.]{
\parbox[h]{\textwidth}
{Raul E. Angulo$^{1,2} \thanks{reangulo@stanford.edu}$, 
Oliver Hahn$^{3,1}\thanks{hahn@phys.ethz.ch}$, 
Tom Abel$^{1}\thanks{tabel@stanford.edu}$.} \vspace*{6pt} \\
\\
$^1$  Kavli Institute for Particle Astrophysics and Cosmology,\\
      Stanford University, SLAC National Accelerator Laboratory, Menlo Park, CA 94025, USA \\
$^2$  Max-Planck-Institut f\"ur Astrophysik, Karl-Schwarzschild-Strasse 1, 85740 Garching bei M\"unchen, Germany. \\
$^3$  Institute for Astronomy, ETH Zurich, CH-8093 Z\"urich, Switzerland.}
\maketitle

\date{\today}
\pagerange{\pageref{firstpage}--\pageref{lastpage}} \pubyear{2013}
\label{firstpage}

\begin{abstract} 
We investigate the large-scale clustering and gravitational interaction of
baryons and dark matter (DM) over cosmic time using a set of collisionless
N-body simulations. Both components, baryons and DM, are evolved from distinct
primordial density and velocity power spectra as predicted by early-universe
physics. We first demonstrate that such two-component simulations require an
unconventional match between force and mass resolution (i.e. force softening on
at least the mean particle separation scale). Otherwise, the growth on {\em
any} scale is not correctly recovered because of a spurious coupling between
the two species at the smallest scales.  With these simulations, we then demonstrate
how the primordial differences in the clustering of baryons and DM are
progressively diminished over time. In particular, we explicitly show how the
BAO signature is damped in the spatial distribution of baryons and imprinted in
that of DM. This is a rapid process, yet it is still not fully completed at low
redshifts. On large scales, the overall shape of the correlation function of
baryons and DM differs by $\sim2\%$ at $z=9$ and by $0.2\%$ at $z=0$. The
differences in the amplitude of the BAO peak are approximately a factor of $5$
larger: $10\%$ at $z=9$ and $1\%$ at $z=0$.  These discrepancies are, however,
smaller than effects expected to be introduced by galaxy formation physics in
both the shape of the power spectrum and in the BAO peak, and are thus unlikely
to be detected given the precision of the next generation of galaxy surveys.
Hence, our results validate the standard practice of modelling the observed
galaxy distribution using predictions for the total mass clustering in the
Universe.
\end{abstract}

\section{Introduction} 

In the standard paradigm of cosmological structure formation, primordial
density perturbations are a result of quantum fluctuations amplified by cosmic
inflation. At these very early times, baryons and dark matter (DM) density
fields have the same phases and amplitudes -- each of them has a fluctuation
spectrum following a power law with an index close to unity. However,
subsequent interaction with the radiation field breaks the initial similarity,
creating a scale-dependent growth that is different for baryons and for DM.
 
On scales smaller than the horizon and prior to recombination, baryons couple
to photons through Compton scattering. Radiation pressure opposes gravity and
inhibits the growth of density perturbations. The balance is not perfect
however, thus generating oscillations on sub-horizon scales in the density,
temperature and size of perturbations in the baryon-photon fluid. On even
smaller scales, free-streaming and an imperfect coupling between baryons and
photons progressively damps the amplitude of these oscillations.  In contrast
to baryons, DM particles do not directly interact with photons, and are thus
mainly affected by gravity. DM density fluctuations can grow freely, and are
only halted by the Meszaros effect on scales smaller than the horizon at the
matter-radiation equality. The physics describing these interactions is
understood at high precision and is able to describe at very high accuracy the
patterns of temperature fluctuations observed in the cosmic microwave
background radiation \citep[see e.g.][for a review]{HuDodelson2002}.

After recombination, baryons decouple from the photons leading to a drop in
sound speed by $\sim5$ orders of magnitude and an associated drop in the Jeans
mass of $\sim14$ orders of magnitude. From now on, the evolution of
perturbations in baryons and DM is dominated by the same physics and the
growth is almost entirely determined by gravity until much later times (when
hydrodynamical interactions become important). The initial conditions for the
two components after recombination are, however, very different. The coupling
between baryons and photons has prevented the baryons from falling into the
density fluctuations present in the DM fluid. The power spectrum of density
fluctuations in baryons and DM are thus genuinely distinct: for instance,
baryonic acoustic oscillations (BAO) dominate the baryon density power spectrum
but they are almost non-existent in the DM distribution.

At later times, the gravitational coupling between DM and baryons will reduce
such differences. The DM distribution gradually obtains a BAO signal, while the
amplitude of BAOs in the baryons clustering is reduced. This process is
commonly assumed to be finished at low redshift, i.e. baryons and  DM
are assumed to have identical spatial distributions (equal to that of the total
mass field) on large scales. A natural corollary of this is that the BAO should be
detectable in the galaxy distribution. This has indeed been achieved
observationally with increasing accuracy
\citep[e.g.][]{Cole2005,Eisenstein2005,Blake2011, Beutler2011} and further
measurements have been proposed using virtually any known tracer of the matter
density field in the Universe: including the galaxy distribution
\citep[e.g.][]{Cooray2002}, galaxy clusters \citep[e.g][]{Angulo2005}, the
$Ly-\alpha$ forest \citep[e.g.][]{White2010,Kitaura2012} or $21$~cm emission
background from galaxies at low redshifts \citep{Chang2008}, from the epoch of
re-ionisation at $z\sim10$ \citep{Mao2008,Rhook2009}, and even using Supernovae
\citep{Zhan2008}.

However, the details of the process in which the DM (and thus the total matter)
distribution acquires the BAO signature still remain relatively unexplored. At
high redshift or large scales, the interaction between baryons and DM particles
can be followed accurately by perturbation theory
\citep{Somogyi2010,Bernardeau2012}.  However, to explore low redshifts and
small scales in the mass field, N-body simulations are essential, since they
provide the most accurate and faithful predictions in the nonlinear regime
\citep[see][for a recent review]{Kuhlen2012}. Unfortunately, to our knowledge,
no N-body simulation has been performed to address this topic. This is an
important issue, since a precise understanding of the BAO signal in the
$z\lesssim 10$ Universe is required to interpret accurately the high precision
measurements that will be carried out over the next decade. 

In this paper, we directly follow the gravitational interaction of DM particles
and baryons, from $z=130$ up to the present day. For this, we perform N-body
simulations of two interacting fluids with different primordial density and
velocity fluctuations: one representing the DM field, and another representing
the baryons. We adopt a canonical cosmological model $\Omega_m=0.276$,
$\Omega_{\Lambda}=0.724$, $\Omega_b = 0.045$, $h = 0.703$, $\sigma_8=0.811$, 
$n_s = 0.961$ \citep{Komatsu2011}. We provide details of these simulations and the numerical set-up
in \S2.  With these simulations in hand, in \S3, we explore the evolution of
the large-scale clustering of baryons and DM, with particular emphasis on the
evolution of the BAO peak. We discuss our results and conclude in \S4.

\section{N-body Simulations}
\subsection{Initial conditions}

We generate the initial position and velocity for our simulation particles at
$z=130$ using the {\sc Music} code \citep{HahnAbel2011}, and adopting a set of
cosmological parameters consistent with the published measurements of the WMAP7
data release \citep{Komatsu2011}. Explicitly: $\Omega_m=0.276$,
$\Omega_{\Lambda}=0.724$, $\Omega_b = 0.045$, $h = 0.703$, $\sigma_8=0.811$ and
spectral index $n_s = 0.961$. 

We compute the primordial power spectra for baryons and DM using a linear
Boltzmann solver code similar to that of \cite{MaBertschinger1995} where
residual baryon-radiation interaction effects become small.  The velocity field
is irrotational and thus fully described by the velocity divergence, so that
the initial conditions are fully specified by the power spectra of the
overdensity $\delta$ and the velocity divergence $\theta$: $P_{\delta_C}$,
$P_{\theta_C}$, $P_{\delta_B}$, $P_{\theta_B}$, where the subscript 'C' stands
for CDM and 'B' for baryons.

These power spectra are used together with the Zel'dovich approximation
\citep[ZA, i.e.  first order Lagrangian Perturbation Theory][]{Zeldovich1970} in
{\sc Music} to generate the initial conditions of our simulations. The 
gravitational potential,
whose gradient appears in both the particles displacement and velocity, is
herein replaced by four potentials generated by the respective density power
spectra and velocity divergence spectra. Hence, the initial
positions and velocities of, for example, baryons is give by:

\begin{equation}
\mathbf{x}_B = \mathbf{q} + \nabla \Phi_B,\quad
\mathbf{v}_B = \nabla \Psi_B,
\end{equation}

\noindent where $\mathbf{q}$ is the Lagrangian coordinate and

\begin{eqnarray}
\Phi_B(\mathbf{k}) &\propto& \mathcal{G}(\mathbf{k})\, k^{-2} \sqrt{P_{\delta_B}(k,t)}, \quad\textrm{and} \\
\Psi_B(\mathbf{k}) &\propto& \mathcal{G}(\mathbf{k})\, k^{-2} \sqrt{P_{\theta_B}(k,t)}
\end{eqnarray}

\noindent are the respective potentials and $\mathcal{G}$ is a real-valued
Gaussian random field of zero mean and unit variance. In this formulation
the streaming velocity between baryons and DM is included
self-consistently, but not its non-linear impact onto the spectra until
$z=130$, which however is negligibly small on the large scales we
consider. 

Due to the simplicity and unambiguity in the formulation of two-fluids initial 
conditions within the ZA, we refrain here from employing the more accurate 
2LPT (2nd order Lagrangian Perturbation Theory) formalism \citep{Scoccimarro1998}.
In addition, we expect the artefacts and inaccuracies introduced by the ZA to have little 
impact in our results. This in part because the high starting redshift of our 
runs, which makes transient features introduced by the ZA have longer time 
to decay in amplitude. And also because we will be mostly concerned in fractional 
differences in the measured clustering, thus small artefacts in the clustering
partially cancel out. 

We also note that at a redshift of $130$ there remains a small contribution
of the radiation energy density to the Friedmann equation. We ignore this
in our simulations and linear theory calculations, setting $\Omega_r=0$ at $z\leq130$,
but not before. 

The baryonic particles are placed on a staggered initial mesh with respect to
the DM particles. The required phase shift of the noise field is computed in
Fourier space. Staggering is necessary to minimise a spuriously tight coupling
between the two particle types \citep[cf.][but also our discussion in
Section~\ref{sec:softening}]{Yoshida2003}.  We also note that the initial
conditions for all our simulations use the same Gaussian white noise field, and
thus the resulting cosmic density fields have identical phases. This allows a 
more accurate comparison between different runs because the effect of cosmic 
variance is largely reduced.

\subsection{Specifics of the N-body simulations}

We use two sets of cosmological N-body simulations to study the gravitational
coupling of baryons and DM. In the first set, all matter is represented by a
single fluid sampling the total matter power spectrum. The second set contains
simulations with two distinct fluids, one representing baryons and the other
DM, both of which have different primordial density and velocity fluctuations,
as predicted by early universe physics. Each set consists of two simulations
with different box sizes: i)~$L = 1\,\Gpc$, sufficient to measure reliably the
BAO signal; and ii)~$L=250\,\Mpc$, with which we focus on smaller scales and on
the coupling between baryons and DM in the nonlinear regime. In all cases, the
DM and baryonic field was each represented with $1024^3$ particles, which for
the $1\,\Gpc$ box implies a DM particle mass of $5.97\times10^{10}\,\Mass$ and
a baryon particle mass of $1.16\times10^{10}\,\Mass$. For the smaller
simulation box these values are $64$ times smaller:  $9.3\times10^{8}\,\Mass$ and
$1.81\times10^{8}\,\Mass$, respectively. Additionally, for the tests presented
in \S2.3 we will employ another set of simulations of mass resolution identical to the $1\,\Gpc$ run, but on a $L=500\,\Mpc$ box. 

We follow the non-linear evolution of the two fluids using a memory-efficient
version of the {\tt P-Gadget3} Tree-PM code \citep{Springel2005a} described in
\cite{Angulo2012}. We compute only gravitational interactions, neglecting all
hydrodynamical ones, thus baryons behave as a collisionless fluid. We expect 
this to be a good first approximation on the large scales and for 
the processes we explore here (this approximation is also adopted in
perturbation theory). While there are works studying how cooling, star formation
and feedback affect the matter power spectrum \citep[e.g.][]{vanDaalen2011}, there 
is yet to be investigated how these processes could affect the detectability of 
BAOs and the coupling between baryons and DM. We regard our work to be the first 
step in that direction.

We would like to highlight that the force resolution of our runs is a critical
issue. A force resolution too high for a given mass resolution 
causes a spurious coupling between DM and baryons, and their respective 
clustering becomes severely affected, departing from the expected value.\footnote{Interestingly, we note that the total mass power spectrum always 
displays the correct amplitude and evolution, independently whether separately
the DM or baryons have the correct power spectrum or not. Both components seem to 
collude to preserve the total mass power spectrum.} 
As we will show in the next subsection, this 
spurious coupling disappears only when the forces are softened on scales larger 
than the mean interparticle separation. For this reason, the main results 
of this paper will be based on simulations where forces are computed only 
with a PM method with a mesh of dimension equal to that of the unperturbed
particle mesh, i.e. using a grid of $1024^3$ points to solve the Poisson 
equation. We present tests in this regard in the next subsection.

\begin{figure} 
\includegraphics[width=8.5cm]{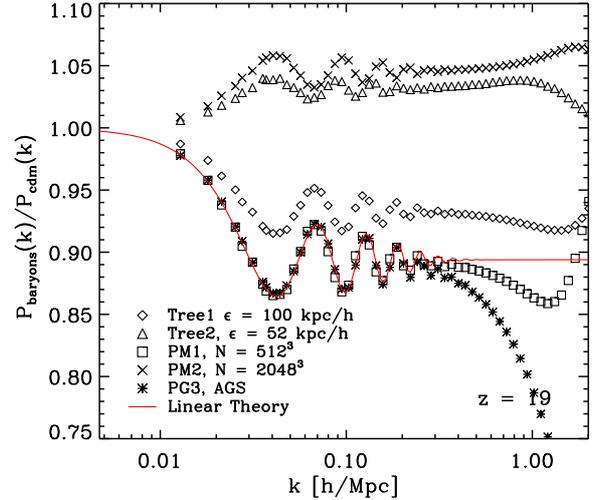} 
\caption{
A test of the effect of force resolution in N-body simulations. The plot
compares the ratio of the power spectrum of baryons and CDM at $z=19$ for four
runs with different force resolution. Triangle and diamonds show results for
TreePm runs with two different softening lengths, $\epsilon$, corresponding to 
$100\,\kpc$ and $52\,\kpc$, respectively.
Squares and crosses show runs where forces are computed
using only a particle-mesh (PM) method with cell sizes equal to
$\sim4\,\Mpc$ and $\sim1\,\Mpc$ (corresponding to $1$ and $0.25$ times the mean
inter-particle separation). Asterisks correspond to a Tree-PM run where
the force acting on every particle is softened adaptively using a SPH kernel whose size is set by the distance to the 32nd nearest neighbour. Finally, the red solid line displays the expectation of linear perturbation theory. \label{fig:force}} 
\end{figure}

\subsection{Force softening and discreteness effects}
\label{sec:softening}

We now present the results of a suite of test runs, which serves as an 
illustration of the spurious coupling between baryons and DM discussed above. 
Fig.~\ref{fig:force} shows the relative difference in the clustering of baryons 
and DM at $z=19$, as predicted by our $L=500\,\Mpc$ simulations. We plot results
from different runs featuring different force resolutions and methods to compute
the gravitational forces. The expectation is set by linear perturbation theory 
and it is displayed by a solid red line. 

Test runs denoted as ``Tree1'' and ``Tree2'' (diamonds and triangles, respectively) 
show results for a standard Tree-PM force calculation and with Plummer-equivalent 
softening length, $\epsilon$, set to $1/20$-th and $1/10$-th of the mean 
inter-particle separation. Despite these being values adopted by state-of-the-art
simulations, in our case they overestimate the strength of the coupling between 
our two particle species providing results completely inconsistent with linear 
theory. 

The same occurs in another test case, denoted as ``PM2'' (crosses), where we 
computed forces using only the PM method where the cell size is equal to 
$1/4$-th of the mean inter-particle separation of each particle type. As in 
the previous test cases, the ratio of the power spectra appears to be 
significantly higher than the prediction of linear theory. Although not
displayed in this figure, we note that these simulations depart even 
further from linear theory at lower redshifts. A similar behaviour of
incorrect large-scale growth has been reported by \cite{Oleary2012}.

We also show the results of another run, denoted as ``AGS'' (asterisks), where forces between baryon and DM particles (not however the intra-species forces) are 
softened adaptively using an SPH kernel with a width set by the distance to
the 32nd neighbour \citep[see Apendix A of][for a discussion of this standard
feature of {\tt Gadget}]{Springel2001a}. Unlike the previous runs, this one 
seems to correctly recover the relative large-scale clustering of baryons and DM. However, this comes at the expense of a strong suppression of the growth of small scales, resulting from a large smoothing of the force field (at high redshifts, 
our SPH kernel size corresponds to $32^{1/3} \sim 3$ times the mean-interparticle
separation). We note that the use of an adaptive softening length
\citep[e.g.][]{Iannuzzi2011} should behave in a similar manner. Also, adaptive softening is inherent in
all cosmological adaptive mesh refinement codes such as {\tt Ramses}
\citep{Teyssier:2002} and {\tt ENZO} \citep{Bryan:1997, OShea:2004}, so that a
spurious coupling between the two species is not expected to occur. For these
codes, the numerical evolution of the baryons on a grid comes however at the
expense of numerical diffusion at the grid scale, leading to a comparable
suppression of high-$k$ modes in the baryons \citep[see Figure~23 of][]{HahnAbel2011} as those in the adaptive SPH run.

Finally, we can see that the ``PM1'' run (squares) also reproduced correctly
the linear growth. In this particular set-up the force has been
smoothed on scales below the mean inter-particle separation for each component
separately (or $1.25$ times the inter-particle separation for both components
together). Here, the agreement with linear theory extends up to $k \sim 0.2\hMpc$.
On smaller scales, we see a downturn caused by the smoothing in the force
field introduced by the Fourier mesh. Nevertheless, this smoothing is not as
strong as that seen in the ``AGS'' case, and reaches the correct behaviour up 
to smaller scales. For 
this reason, we will adopt this particular numerical configuration for the simulations used in the remainder of the paper.

We can track the origin of the spurious coupling to the collissionality 
appearing in runs with high force
resolution as a result of the discretisation of the density fields. On
small scales, the force generated by N-body particles only approximately
represents an homogeneous and continuous force field. This fact causes the
particles to quickly couple on small scales, which affects the evolution even
on large scales, increasing the rate at which differences in the clustering of
baryons and DM dissipate. 

We also note that we did not observe a significant impact of the choice of
initial particle distribution on large-scale behaviour. While
\cite{Yoshida2003} and \cite{Naoz2011}, e.g., have argued for two randomly
displaced glass distributions for the two species, we find that any choice of
`staggered' initial distribution -- where separations between particles from
the two species are locally maximized -- reduces, but does not avoid, the
discreteness effects. Regardless of the initial particle distributions, a very
large force softening has to be chosen to obtain the correct growth.  The
reason why \cite{Yoshida2003} \citep[and][]{HahnAbel2011} do not observe a
similar spurious coupling is most likely due to the significantly smaller boxes
investigated there, where strong non-linear growth dominates quickly over the
numerical artefacts.

It is important to note that the reason why discreteness effects in standard 
one- or two-component particle simulations are not as evident as in our case
\citep[e.g.][]{Hamana2002} is because these simulations typically start from
identical perturbation spectra, so that the spurious coupling cannot be easily
diagnosed using the power spectrum. We argue, however, that in two-component
simulations (such as SPH+DM simulations), it should still appear as an
additional binding energy between the two fluids that has to be overcome by
pressure forces at late times. The impact is even less clear though, and
harder to quantify, in standard one-fluid CDM calculations, but we expect it to
appear as an artificial population of small-scale of low-mass halos, somewhat
similar to those seen in warm DM simulations \citep[e.g.][]{Wang2007}. Whether
this has any sizeable and unforeseeable consequence for the results of
simulations remains to be investigated.

\section{The Clustering of Baryons and Dark Matter}

Having identified a robust numerical setup, in this section, we present
predictions for the large-scale clustering of baryons and dark matter as
measured in N-body simulations. We start showing results in Fourier space
(\S3.1), then move to configuration space where we center our discussion on the
BAO peak (\S3.2).

\subsection{Power Spectrum}

\begin{figure*} 
\includegraphics[width=15cm]{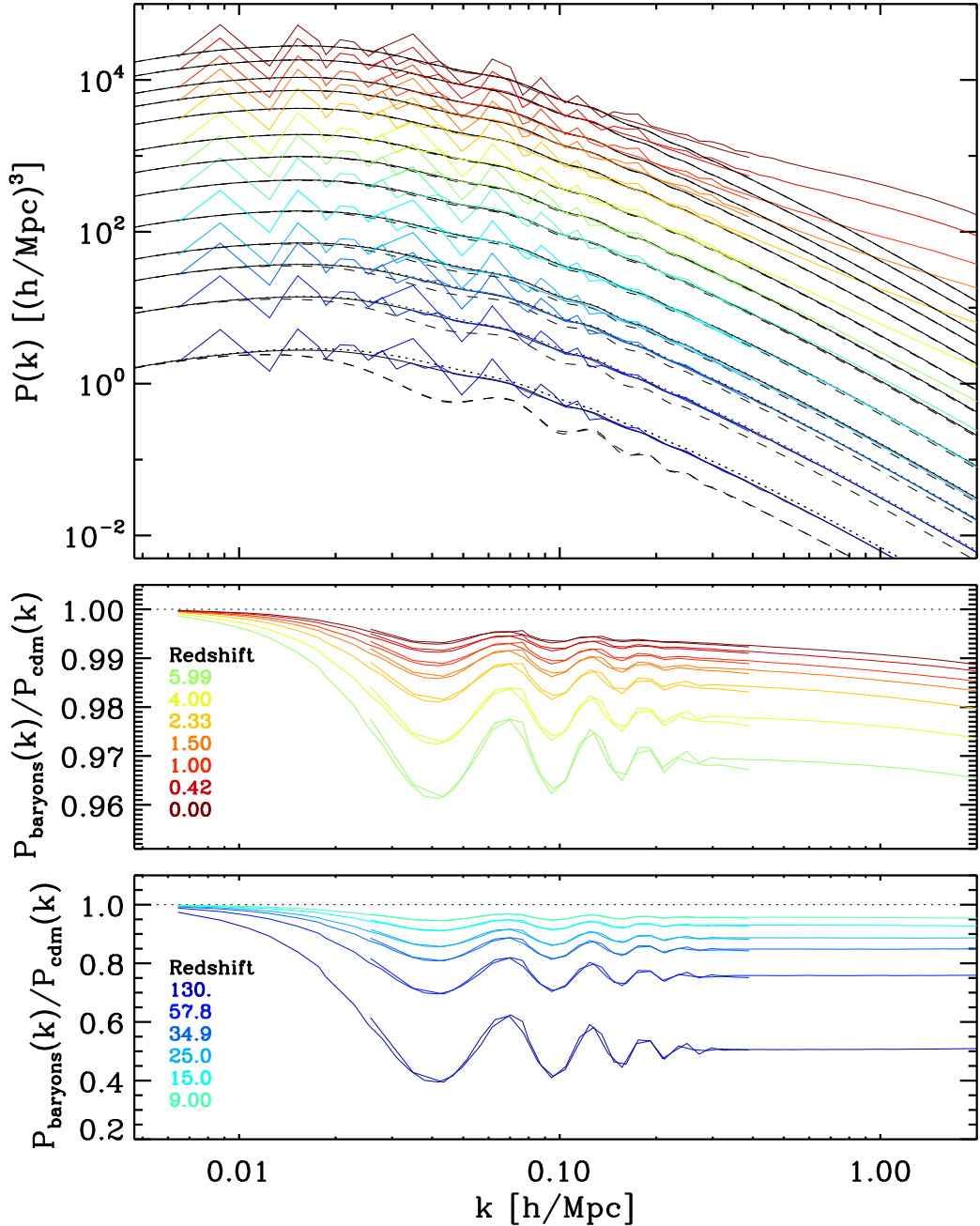} 
\caption{
The growth of density fluctuations in the baryon and in dark matter fields, as
measured by the power spectrum in real space. We show
results for $13$ redshifts, as indicated by the annotations in the figure, and
for two different simulations containing the same number of particles but on
boxes of side $1000\,\Mpc$ or $250\,\Mpc$. Their predictions overlap over the
range $k = [0.03 - 0.3] \hMpc$.  In the {\bf top panel}, coloured lines show
the measured total mass power spectrum in our simulations whereas dashed,
dotted and solid lines show predictions of linear perturbation theory for the
baryon, CDM and total mass power spectra, respectively.  The {\bf lower panels}
show the ratio between the measured power spectra from baryons to that from
CDM.  Note that this ratio is not affected by the noise that arises from the
finite sampling of large modes present in the simulation box. No Poisson
shot-noise has been subtracted.  Note the suppression of baryon perturbations
relative to CDM perturbations at large $k$ that is due to non-linear effects.
\label{fig:pks}} 
\end{figure*}

\begin{figure} 
\includegraphics[width=8.7cm]{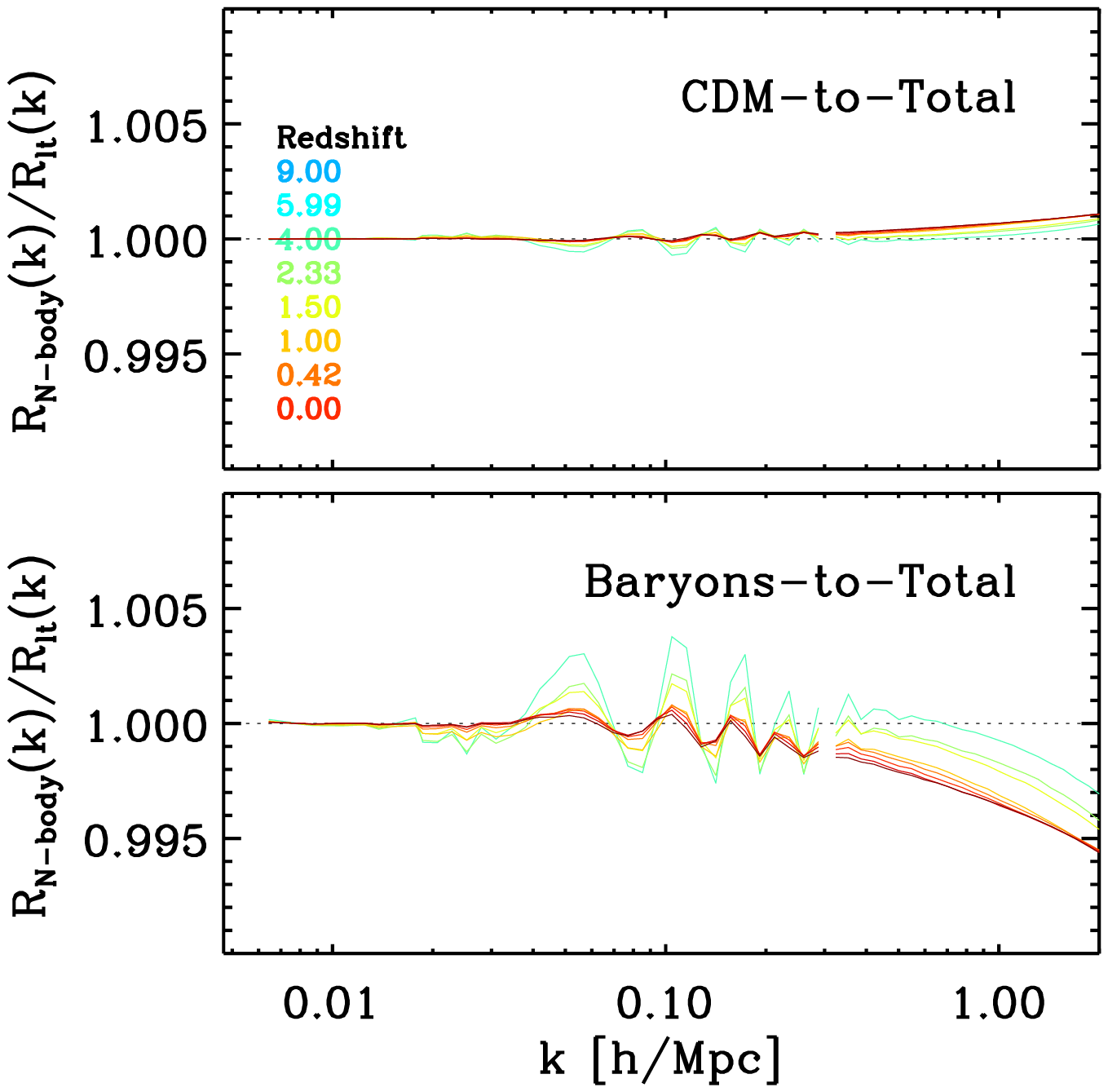} 
\caption{ Comparison between simulation results, $R_{N-body}$, and 
linear perturbation theory predictions, $R_{lt}$, for the relative 
difference  between the power spectrum of baryons or CDM (top and bottom panel, 
respectively) to the total mass field: $R(k)=P_{b,cdm}/P_{total}$.
We show results from our $1\Gpc$ box for $k < 0.3\,\hMpc$, and from our
$250\,\Mpc$ for smaller scales. Coloured lines show results at different
redshifts, as indicated by the legend.
\label{fig:lt}} 
\end{figure}

We begin by presenting in Fig.~\ref{fig:pks} the power spectra, P(k), of DM and
baryons in real space, measured from our simulations at different output
redshifts.  Measurements were performed by mapping the particle distribution,
using a clouds-in-cell scheme, onto a $1024^3$ grid and then Fast Fourier
Transforming this field. We correct the effects of the assignment scheme by
dividing each mode by the Fourier transform of a cubical top-hat, but we do not
subtract a Poisson shot-noise term.  Here we show results from our large- and
small-box simulations.  The minimum wavenumber plotted is $2\pi/1000 =
0.0062\,\hMpc$ and $2\pi/250 = 0.025\,\hMpc$ for each of them.

The top panel of Fig.~\ref{fig:pks} compares the growth of total mass power
spectrum from our runs with that predicted by linear perturbation theory, which
is displayed as solid black lines. We note that we have evolved the $z=130$
linear theory power spectrum without considering the interaction of photons
and baryons (effectively setting $\Omega_r = 0$), which allows a more direct
comparison with our simulations, where only gravitational interactions are
considered.

We can see that our simulations closely reproduce the expected linear growth,
differing only at the $4\%$ level: The growth of the fundamental mode, from
$z=130$ until $z=0$, is $10438.5$, whereas the linear theory prediction is
$10015.6$.  This close agreement supports the correctness of our numerical
calculation. On small scales, the expected nonlinear growth dominates, and the
measured $z=0$ power spectrum is a factor of $\sim20$ larger than linear theory
expectations at $k = 1\,\hMpc$. 

In the two bottom panels of Fig.~\ref{fig:pks} we plot the ratio between the
power spectra of baryons and DM. This highlights the differences in the overall
shape as well as in the BAO signature, visible in the range $0.1 < (k/\hMpc) <
0.3$.  At all redshifts, the curves are systematically different from unity,
which implies that the overall shape of the power spectrum of baryons and DM is
different, even on very large scales. Density perturbations in the baryon
density field are smaller than those in the DM at all times, resulting from the
extra suppression produced by radiation pressure before recombination.  

At the starting redshift of our simulations ($z=130$), the amplitude of the
power spectrum of baryons is $40\%$ of that of the DM component, even on scales
as large as the turn-over ($k=0.02\,\hMpc$). On smaller scales, the difference
is constant down to the smallest scales that our simulations resolve.  At later
times, the gravitational interaction between baryons and DM particles couples
the perturbation fields at all scales. As a result, both fields are homogenised
and initial differences are progressively reduced. At $z=35$ the baryon to DM
P(k) ratio is $\sim0.85\%$, and $\sim0.95\%$ at $z=9$. At the latter redshift,
fluctuations have been amplified by a factor of $100$, but the differences
between the baryon and DM power spectra are still approximately scale
independent. 

Gravitational interaction continues at lower redshifts, but, due to the effect
of dark energy, perturbations grow more slowly than at higher redshifts. By
$z=0$ the baryons power spectra show a $1\%$ suppression compared to that of
DM, almost the same difference present at $z=2$. The ratio now displays a scale
dependence -- small scales approach unity at a slower rate -- which can be
interpreted as the DM field having experienced more nonlinear evolution and
mode coupling than the baryonic field. This could be a residual effect of the
DM power spectrum having higher amplitude than that of baryons. 

The amplitude of the oscillatory behaviour seen in the lower panels of
Fig.\ref{fig:pks} arises from differences in the BAO feature.  The bigger
their amplitude, the more dissimilar the BAO are in the two fluids. At the
starting redshift, oscillations are large, i.e. BAOs are very weak in the DM
but very notorious in the baryons. (For baryons, the amplitude of BAO is
$\sim30\%$ of that of the power spectrum, but for DM, it is only $\sim5\%$.) At
lower redshifts, gravity will damp the BAO in the baryons and increase their
amplitude in the DM. However, as in the case of the overall shape of the power spectra,
even at $z=0$, there are still some residual differences. In the next
subsection we will investigate the evolution of the shape and amplitude of BAO
signal in more detail. 

Fig.~\ref{fig:lt} shows a detailed comparison of our results with the
prediction of linear perturbation theory. The top/bottom panel shows the ratio
between the power spectrum of DM/baryons to that of the total mass, divided by
the same ratio predicted in linear theory. Thus, departures from unity indicate
where linear theory is not able to predict the relative differences, with
respect to the total mass, present in the spatial distribution of DM or
baryons. Firstly, we note that departures in this ratio are much smaller, and
appear on smaller scales, than absolute deviations of each component with
respect to linear theory. Here, differences are below $0.5\%$ for the baryons
and below $0.1\%$ for the DM, to be compared with one order of magnitude
deviations in their individual amplitude (c.f. Fig.~\ref{fig:pks}). Hence, linear
theory predicts this quantity at a remarkable accuracy.  We note that our
findings qualitatively agree with the analytical results of \cite{Somogyi2010},
who, using renormalised perturbation theory extended to a multi-fluid case,
also reported a positive deviation from unity on small scales for DM, and
negative deviation for baryons. 

Overall, from this figure we get a picture complementary to our earlier
discussion: Nonlinear evolution produces a faster growth in DM perturbations
relative to that in baryons. For this reason, the amplitude of the baryon
spectrum is overestimated, whereas that of DM is underpredicted. This results
in an overprediction of the baryon-to-DM power spectra ratio, which increases
at smaller scales. Linear theory predicts only a $1\%$ difference between the
clustering of DM and baryons by $z=0$, whereas in our simulations we measure a
$1.5\%$ difference. These departures are small, but they should increase as we consider 
smaller scales and, nevertheless, exemplify that gravitational evolution can be 
followed at some degree by linear or higher-order perturbation theories, but once 
the density contrast reaches values well above unity, numerical simulations are 
needed to properly and accurately follow the growth of structure.  

\subsection{The BAO peak in the Correlation Function}
 
\begin{figure*} 
\includegraphics[width=17.5cm]{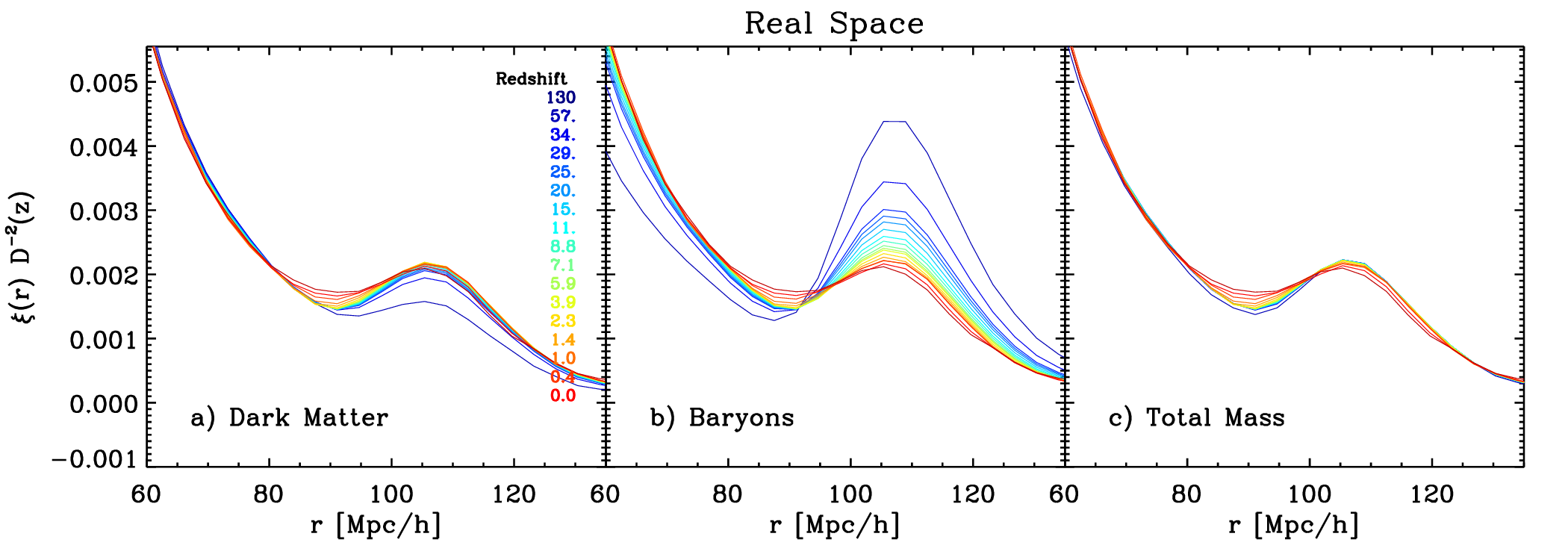} 
\includegraphics[width=17.5cm]{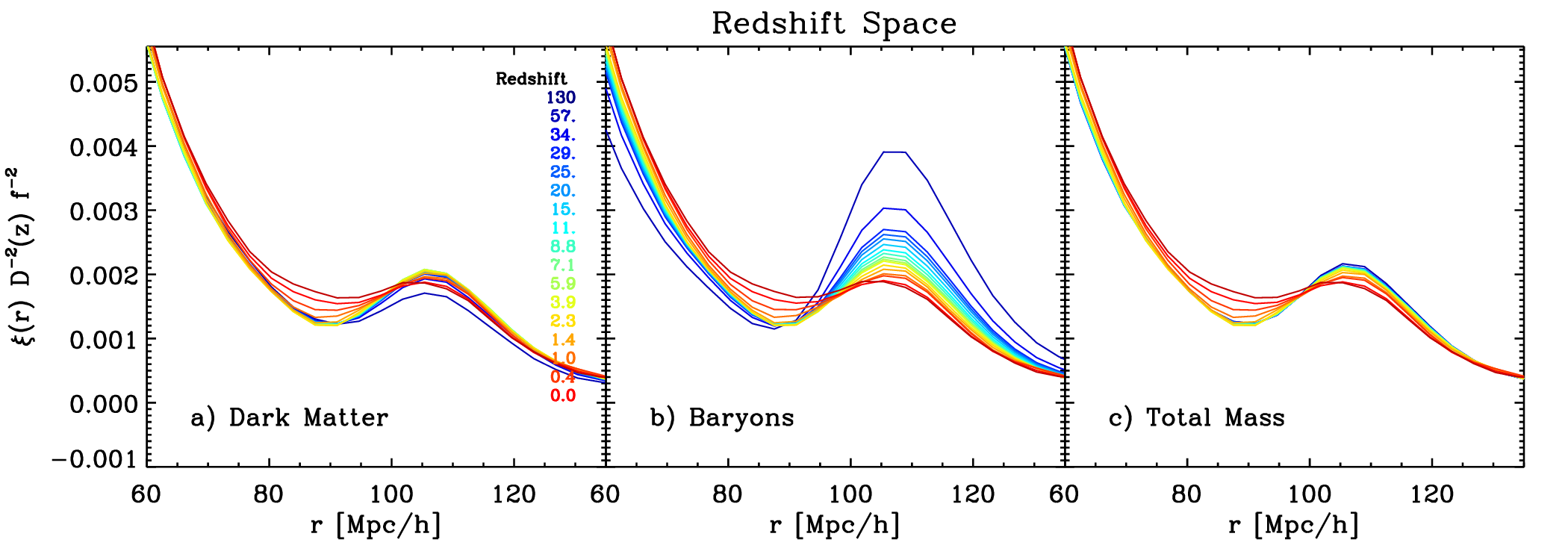} 
\caption{Real-space (top) and redshift-space (bottom) correlation functions in our
simulations at different redshifts, focusing in the BAO peak. Measurements are
shown for baryons, dark matter and total mass (panels (a), (b) and (c)), and
for linear bins of $\Delta r = 3\,\Mpc$. Results were scaled by the square of
the appropriate growth factor, $D(z)$, and, in the case of the redshift-space correlation functions, additionally by the Kaiser boost factor: 
$f \equiv 1 + \frac{2}{3}\beta + \frac{1}{5} \beta^2$, where 
$\beta \approx \Omega(z)^{0.55}$. \label{fig:bao_r}} 
\end{figure*}

\begin{figure} 
\includegraphics[width=8.5cm]{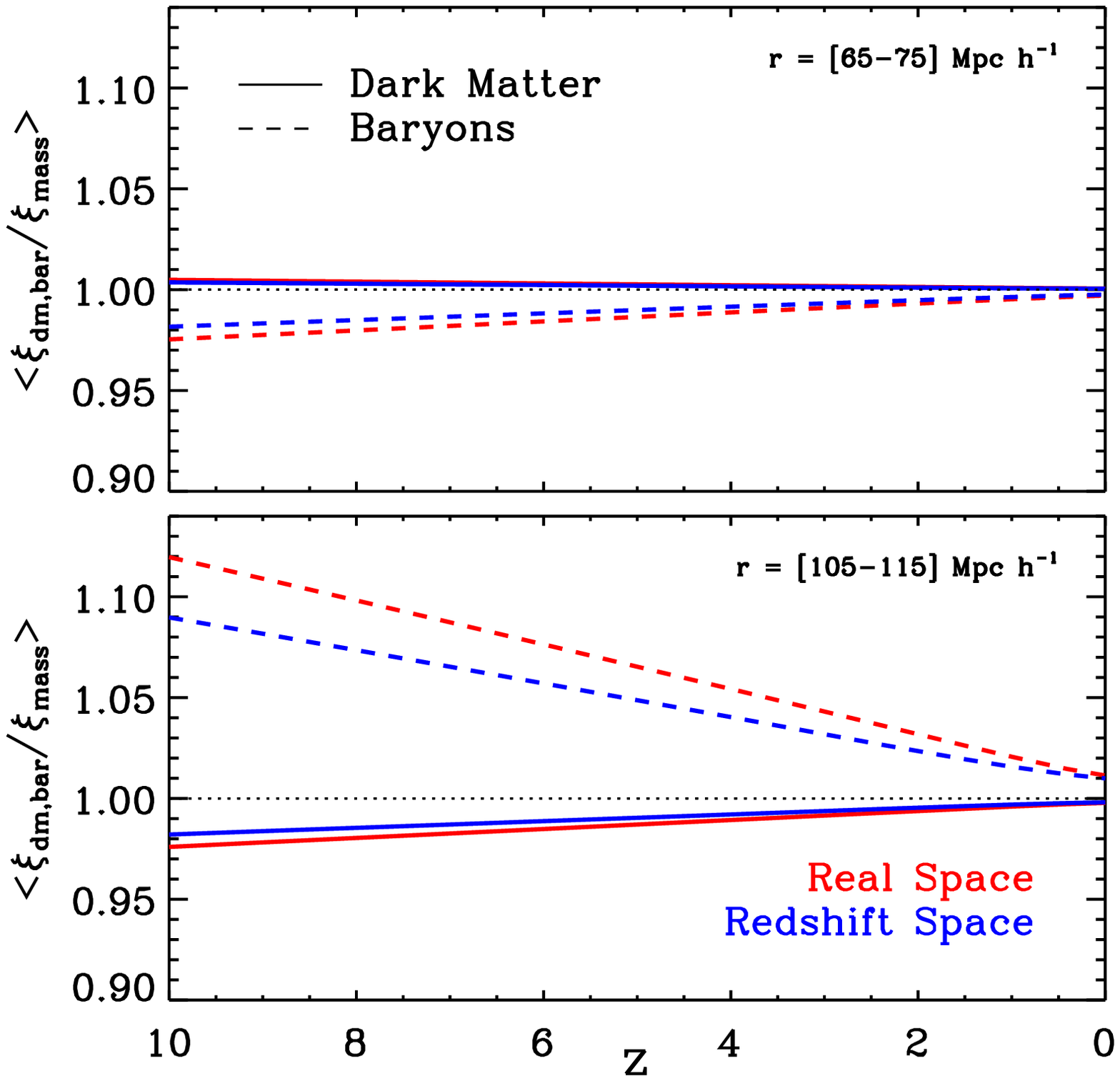} 
\caption{
The redshift evolution of the average correlation function for baryons and DM relative
to that of the total mass. In the top panel, we show the ratio of the
correlation functions averaged between $65$ and $75\,\Mpc$ separations. Thus,
this displays differences in the overall amplitude of the respective
correlation function. In the bottom panel, we perform the average between $105$
and $110\,\Mpc$. Thus this can be regarded as the evolution in the amplitude of
the BAO peak for baryons and DM.  \label{fig:dif}} 
\end{figure}

We now complement the Fourier-space results by presenting the correlation
function, $\xi(r)$, of baryons, DM and that of the total mass, as measured in
our simulations. Exploring our results in configuration space allows us
to focus on the BAO and also to have a more direct comparison with observational data and expectations from future galaxy surveys, which usually present their
measurements in the form of correlation functions. 

We compute correlation functions in Fourier space using a
mesh of $1024$ points per dimension, which provides an accurate estimation of
correlation functions on $r \gtrsim 20\,\Mpc$ for our $1\,\Gpc$ box.  This
approach is considerably faster than a direct pair count when computing
correlation functions on large scales in catalogues containing a large number
of particles.

Fig.~\ref{fig:bao_r} shows the correlation functions computed in this way for
both baryons and DM in our $1\,\Gpc$ simulation. The top panel displays the
measurements in real space while the bottom panel shows the results in redshift
space using a plane-parallel approximation (i.e. the effect of peculiar motions
in distance estimators is included by considering an observer at infinity).  We
show results on scales $r = [60-135]\Mpc$ to focus on the BAO peak (appearing
at $\sim110\,\Mpc$).  Each measurement has been divided by the square of the
growth factor, and additionally by the Kaiser ``boost factor"
\citep{Kaiser1987}, in the case of redshift-space results. Therefore, if the
growth of a given component is completely linear and scale-independent, then
curves at different redshifts would coincide in each panel. We see that this is
indeed a good first-order approximation, though systematic differences exist.
We discuss this next.

In the top-left panel of Fig.~\ref{fig:bao_r} we can see how the BAO peak is
gradually imprinted in the DM distribution. At $z=130$, the correlation
function is still close to a power-law with just a relatively small bump at
$110\,\Mpc$.  However, in an extremely rapid process, the BAO emerges and
reaches almost its full amplitude by $z\simeq30$, only $\sim100$~Myrs after.
During this period, we also see a small scale-dependent suppression of the
correlation function at scales $r<80\,\Mpc$, to accommodate for the growing
peak. The subsequent evolution happens at a lower rate, and the maximum
relative amplitude of the BAO is reached at $z\sim6$, the moment in which mild
nonlinear coupling of independent Fourier modes start to have a noticeable
effect, decreasing again the amplitude of BAO peak and broadening its shape
\cite[e.g.][]{Angulo2005}.

The history of baryons, displayed in the middle panels, is the opposite.  At
the starting redshift the correlation function is dominated by the BAO peak, it
is $\sim4$ times larger than the broad-band shape of the correlation function,
and a factor of $\sim2$ larger than the maximum relative amplitude ever present
in the total mass field. This, however, quickly changes, with the amplitude of
the peak decreasing linearly with redshift. As in for the DM case, density
fluctuations in the baryons also show a scale-dependent growth, but in the
opposite direction.

The evolution in baryons and DM exactly compensate each other to give an almost
perfectly scale-independent growth of fluctuations in the total mass density,
as can be seen in the right panels of  Fig.~\ref{fig:bao_r}.  This behaviour is
only broken at low redshift by non-linear mode coupling, which smears out the
BAO peak \citep[e.g.][and references therein]{Angulo2008,Sanchez2008}.  The
evolution is compensated since perturbations do not grow independently, but are
linked through a common gravitational potential that determines an identical
acceleration field for both components. Thus, gravity acts as a homogenizer of
the fields. In fact, although by $z=0$ there are residual differences, if we
let the simulation run into the future, then eventually both fields will be
indistinguishable from each other on large scales.

The bottom panel of Fig.~\ref{fig:bao_r} shows the clustering in {\it redshift 
space} (a quantity closer to that observed by spectroscopic galaxy surveys) for 
baryons, DM and for the total mass. For both
components, and at all redshifts, we can see that the measured correlation
functions are not simply a scaled version of their real-space counterpart, as
one might naively expect in linear theory. 

At high redshift, the redshift-space enhancement in the overall correlation
function is smaller than linear theory expectations for DM, but larger for
baryons. At low redshifts, predictions are closer to our measurements, and the
respective curves therefore overlap better in this plot.  At the same time, the
BAO peaks get damped as a consequence of nonlinear contributions to peculiar
velocities.  At high redshifts, this is also true for baryons but not for the
DM field, which sees an increase in the amplitude of the BAO peak.
 
All the differences described above are summarised and quantified in
Fig.~\ref{fig:dif}, which displays the correlation function for baryons and DM
relative to that of the mass for different redshifts and for both real and
redshift space, averaged over two scales. In the top panel, averaged between
$65$ and $75\,\Mpc$, capturing changes in the overall amplitude. In the bottom
panel, between $105$ and $110\,\Mpc$, so it shows changes in the amplitude of
the BAO peak.
 
We see that the correlation function amplitude at $z=10$, relative to that of
the total mass, is $\sim1\%$ higher for the case of DM, and $\sim3\%$ smaller
for baryons.  These figures are reduced to a sub-percent level by $z=0$. The
differences are largest for the BAO peak in the baryons: $\sim10\%$ excess at
$z=10$, which is reduced to about $1$ percent by $z=0$. This effect is smaller
than those introduced by galaxy formation physics (Angulo~et~al. 2013, in
prep.), which could cause up to $\sim10\%$ effects.  

Redshift space distortions (RSD) reduce the amplitude of the BAO peak in the baryons
at all redshfts, relative to that in real space. For DM, they enhance the BAO
peak at high redshifts, but damp it after $z\sim1$.  The observed behaviour  is
explained by the fact that the gravitational potential that originates the
coherent velocity flows is generated by the total mass distribution, which is
not identical to either the baryonic or DM material but to their mass weighted
average. Thus, the density-velocity relation expected in linear theory does not
hold separately for either of the components. At high redshift this homogenises the
field further at the BAO scale, whereas at low redshift nonlinear RSD dominate
and diffuse the peak.
 
\subsection{Comparison with 1-fluid simulation}

\begin{figure} 
\includegraphics[width=8.5cm]{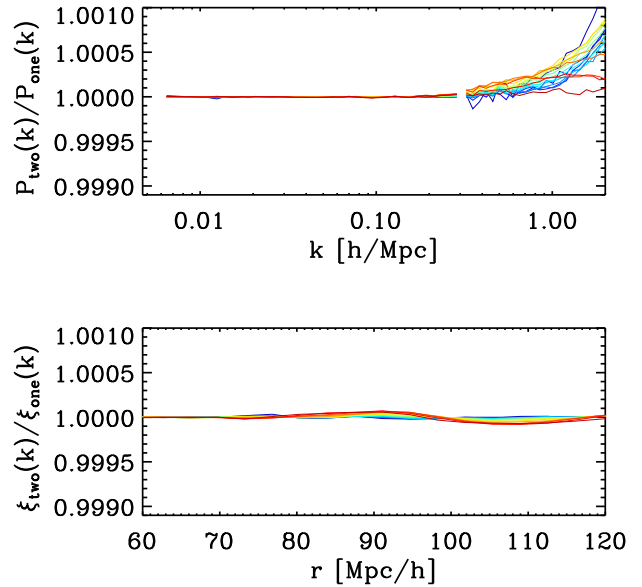} 
\caption{
Comparison of the power spectrum (top panel) and correlation function (bottom
panel) predicted by our two-fluid simulations with those predicted by a
single-fluid simulation, in which the initial conditions of particles were
drawn following the total mass power spectrum at the starting redshift of our
simulations. Lines of different colours display results at different redshifts,
following the same convention as in Fig.~\ref{fig:lt}.
\label{fig:onetwo}} 
\end{figure}

Finally, we test the widespread assumption that the nonlinear evolution of
perturbations in the density field of baryons and DM can be approximated as a
single fluid representing the total mass field.  For this, we run our two
N-body simulations with exactly the same numerical configuration, but this time
baryons and DM have identical primordial density and velocity power spectra
(equal to that of the total mass), so that they effectively behave as a single
fluid. With this we can test to what degree the nonlinear evolution of two
interacting fluids is equivalent to that of a single fluid with the
mass-weighted average power spectrum.

Fig.~\ref{fig:onetwo} presents our results. In the top panel we show the ratio
of the power spectra of our two- and one-fluid simulations. On large scales,
both simulations are virtually indistinguishable, it is only in the nonlinear
regime where differences appear. The one fluid case underestimates the amount
of nonlinear clustering in a roughly redshift-independent manner. The
differences are small: at $k \lesssim 2\hMpc$ are less than $0.1\%$, although
they increase exponentially with the wavenumber $k$ to reach percent level by
$k\sim10$. Unfortunately, our simulations lack the spatial resolution to pin
down this more accurately. Our results agree with the analytical findings of
\cite{Somogyi2010}, who found a discrepancy of less than $\sim0.3\%$ at $k =
1\,\hMpc$ between the mass power spectra from one- and two-fluid calculations
at $z=0$. 

In the bottom panel of Fig.~\ref{fig:onetwo} we display an analogous comparison
but focusing on the BAO peak. Contrary to the Fourier-space behaviour, we see
that the amount of nonlinear diffusion of the BAO peak is larger in the one
fluid case. Nevertheless, the differences are extremely small: below $0.01\%$.
These results validate the use of one-fluid simulations to study the
large-scale distribution of mass, and hence of the BAO signal, in the Universe.
Although not shown in the figure, we have also checked that results in redshift
space present equally small differences.

In this section we focused on two-point statistics, but higher-order 
statistics, in principle, are also be affected. However, we do not expect to find 
differences considerably higher than those shown for the power spectrum and 
correlation function. This, given the very small discrepancies we see between 1 and 
2-fluid simulations in the nonlinear regime, and that these nonlinear interactions
are the responsible for higher order correlations of the density field.

\section{Summary and Conclusions}

We have performed a direct calculation of the gravitational coupling of baryons
and dark matter, from $z=130$ to $z=0$, using N-body simulations. We focused
primarily on large scales to explore how the BAO signature, initially present
primarily only in the spatial distribution of baryons, is imprinted in the DM
distribution. 

At the starting redshift, the density and velocity power spectra of
baryons and DM differ considerably. At latter times, gravity acts as a diffusive
agent of the primordial clustering differences.  We find that the bulk of the
differences are dissipated quickly, but also that this process is not fully
completed by the present day. Differences in the power spectrum
amplitude of $\sim40\%$ at $z=130$ are brought down to sub-percent level by
$z=0$. Interestingly, the BAO peak does not evolve as quickly and differences
are still above $1\%$ at low redshifts. 

On large scales, our results are consistent with linear perturbation theory:
the ratio between the power spectra of baryons and DM decreases in a roughly
scale-independent manner at high redshifts. On small scales,
$k\gtrsim0.2\,h{\rm Mpc}^{-1}$, nonlinear evolution breaks this and we predict
a growth of the ratio between the baryon and the DM power spectrum that is
slower than in linear theory. However, the size of these departures is small (below
$0.5\%$), thus, in practice, linear theory is a very accurate predictor of all
these effects.

Additionally, we confirmed that the nonlinear evolution of the total mass power
spectrum can be accurately predicted by numerical simulations treating all mass
as a single fluid with a single power spectrum. Most of these results are in
qualitative agreement with the analytical work of \cite{Somogyi2010}.

We conclude that assuming that baryons and dark matter have the same spatial
distribution on the BAO scale is a good approximation given the accuracy with
which it is expected to be measured by upcoming large galaxy surveys. However,
we emphasize that the quality of this assumption is redshift-dependent, being
worse at high redshifts and better at low redshifts. The amplitude in the BAO
peaks differs at the $10\%$ level at $z=10$, but only at $1\%$ at $z=0$. The
differences at low redshifts are smaller than the effects expected to be
introduced by galaxy formation physics, but they could perhaps eventually be
detected in the future in, for instance, HI clustering from SKA measurements,
or in another. If this is the case, our work indicates that at high redshift
the bias of tracers that are sensitive to baryons instead of the total matter
distribution might include also a non-negligible baryon bias, though more work
in this respect is needed, since the size of the effect likely depends on the
type of tracer and object selection criteria employed. 

We note that it is certainly possible that the distortions discussed in our 
paper could result into a systematic bias in cosmological parameters 
extracted from BAO measurements. These are, however, hard to predict in practice
due to several reasons. The first one is a theoretical issue, and has to do 
with the question of whether galaxies will follow the baryons, the DM, or the 
total mass. In other words, is the galaxy bias simplest when expressed with 
respect to the underlying DM, baryon or mass clustering? The second issue 
has to do with how the cosmological parameters are actually measured. For 
instance, at low redshifts, distortions in the amplitude are very degenerated 
with the amount of nonlinear evolution in the BAO peak. Thus, amplitude changes
will likely not introduce any biases in the measured sound horizon scale. At 
higher redshifts this is not true, and the effects discussed here should be 
considered In any case, for a quantitative assessment, all the details of a
particular observational campaign should be taken into account. We anticipate
that this will be an interesting area of research in the future.\textsl{•}

Finally, we highlighted that force resolution is a critical issue to obtain
accurate results for simulations with two fluids with distinct primordial
density fluctuations. Discreteness effects in standard numerical configurations
resulted in an artificial coupling between baryons and DM particles, which
propagated to large scales and lead to an artificially fast growth of
large-scale modes. We tracked this issue to the loss of collisionallity in the
fluids, which, thus, could be solved only by softening gravitational forces
below the mean inter-particle separation. The numerical artefacts arising
from a high force resolution were evident in our runs, but they should also be
present in standard N-body calculations. However, while the net effect is not
clear, it clearly warns for a careful assessment of the robustness of current
numerical simulations in the mildly non-linear regime.
 
\section*{Acknowledgements}
REA acknowledges useful discussions with T. Naab, A. Sanchez, V. Springel and
S. White. We would also like to thank Naoki Yoshida, Robert Smith and Simon
White for valuable suggestions and comments on the draft.  OH acknowledges
support from the Swiss National Science Foundation (SNSF) through the Ambizione
fellowship.  TA acknowledges support by the LDRD program at the SLAC National
Accelerator Laboratory as well as the Terman fellowship at Stanford University.
We gratefully acknowledge the support of Stuart Marshall and the SLAC
computational team, as well as the computational resources at SLAC.

\bibliographystyle{mn2e} \bibliography{bao}

\begin{thebibliography}{37}
\expandafter\ifx\csname natexlab\endcsname\relax\def\natexlab#1{#1}\fi

\bibitem[{{Angulo} {et~al}\mbox{.}(2005){Angulo}, {Baugh}, {Frenk}, {Bower},
  {Jenkins}, \& {Morris}}]{Angulo2005}
{Angulo} R., {Baugh} C.~M., {Frenk} C.~S., {Bower} R.~G., {Jenkins} A.,
  {Morris} S.~L., 2005, \mnras, 362, L25

\bibitem[{{Angulo} {et~al}\mbox{.}(2008){Angulo}, {Baugh}, {Frenk}, \&
  {Lacey}}]{Angulo2008}
{Angulo} R.~E., {Baugh} C.~M., {Frenk} C.~S., {Lacey} C.~G., 2008, \mnras, 383,
  755

\bibitem[{{Angulo} {et~al}\mbox{.}(2012){Angulo}, {Springel}, {White},
  {Jenkins}, {Baugh}, \& {Frenk}}]{Angulo2012}
{Angulo} R.~E., {Springel} V., {White} S.~D.~M., {Jenkins} A., {Baugh} C.~M.,
  {Frenk} C.~S., 2012, \mnras, 426, 2046

\bibitem[{{Bernardeau} {et~al}\mbox{.}(2012){Bernardeau}, {van de Rijt}, \&
  {Vernizzi}}]{Bernardeau2012}
{Bernardeau} F., {van de Rijt} N., {Vernizzi} F., 2012, \prd, 85, 063509

\bibitem[{{Beutler} {et~al}\mbox{.}(2011){Beutler}, {Blake}, {Colless},
  {Jones}, {Staveley-Smith}, {Campbell}, {Parker}, {Saunders}, \&
  {Watson}}]{Beutler2011}
{Beutler} F. {et~al.}, 2011, \mnras, 416, 3017

\bibitem[{{Blake} {et~al}\mbox{.}(2011){Blake}, {Kazin}, {Beutler}, {Davis},
  {Parkinson}, {Brough}, {Colless}, {Contreras}, {Couch}, {Croom}, {Croton},
  {Drinkwater}, {Forster}, {Gilbank}, {Gladders}, {Glazebrook}, {Jelliffe},
  {Jurek}, {Li}, {Madore}, {Martin}, {Pimbblet}, {Poole}, {Pracy}, {Sharp},
  {Wisnioski}, {Woods}, {Wyder}, \& {Yee}}]{Blake2011}
{Blake} C. {et~al.}, 2011, \mnras, 418, 1707

\bibitem[{{Bryan} \& {Norman}(1997)}]{Bryan:1997}
{Bryan} G.~L., {Norman} M.~L., 1997, in Astronomical Society of the Pacific
  Conference Series, Vol. 123, Computational Astrophysics; 12th Kingston
  Meeting on Theoretical Astrophysics, {Clarke} D.~A., {West} M.~J., eds., pp.
  363--+

\bibitem[{{Chang} {et~al}\mbox{.}(2008){Chang}, {Pen}, {Peterson}, \&
  {McDonald}}]{Chang2008}
{Chang} T., {Pen} U., {Peterson} J.~B., {McDonald} P., 2008, Physical Review
  Letters, 100, 091303

\bibitem[{{Cole} {et~al}\mbox{.}(2005){Cole}, {Percival}, {Peacock}, {Norberg},
  {Baugh}, {Frenk}, {Baldry}, {Bland-Hawthorn}, {Bridges}, {Cannon}, {Colless},
  {Collins}, {Couch}, {Cross}, {Dalton}, {Eke}, {De Propris}, {Driver},
  {Efstathiou}, {Ellis}, {Glazebrook}, {Jackson}, {Jenkins}, {Lahav}, {Lewis},
  {Lumsden}, {Maddox}, {Madgwick}, {Peterson}, {Sutherland}, \&
  {Taylor}}]{Cole2005}
{Cole} S. {et~al.}, 2005, \mnras, 362, 505

\bibitem[{{Cooray}(2002)}]{Cooray2002}
{Cooray} A., 2002, in Astronomical Society of the Pacific Conference Series,
  Vol. 283, A New Era in Cosmology, {Metcalfe} N., {Shanks} T., eds., p. 162

\bibitem[{{Eisenstein} {et~al}\mbox{.}(2005){Eisenstein}, {Zehavi}, {Hogg},
  {Scoccimarro}, {Blanton}, {Nichol}, {Scranton}, {Seo}, {Tegmark}, {Zheng},
  {Anderson}, {Annis}, {Bahcall}, {Brinkmann}, {Burles}, {Castander},
  {Connolly}, {Csabai}, {Doi}, {Fukugita}, {Frieman}, {Glazebrook}, {Gunn},
  {Hendry}, {Hennessy}, {Ivezi{\'c}}, {Kent}, {Knapp}, {Lin}, {Loh}, {Lupton},
  {Margon}, {McKay}, {Meiksin}, {Munn}, {Pope}, {Richmond}, {Schlegel},
  {Schneider}, {Shimasaku}, {Stoughton}, {Strauss}, {SubbaRao}, {Szalay},
  {Szapudi}, {Tucker}, {Yanny}, \& {York}}]{Eisenstein2005}
{Eisenstein} D.~J. {et~al.}, 2005, \apj, 633, 560

\bibitem[{{Hahn} \& {Abel}(2011)}]{HahnAbel2011}
{Hahn} O., {Abel} T., 2011, \mnras, 415, 2101

\bibitem[{{Hamana} {et~al}\mbox{.}(2002){Hamana}, {Yoshida}, \&
  {Suto}}]{Hamana2002}
{Hamana} T., {Yoshida} N., {Suto} Y., 2002, \apj, 568, 455

\bibitem[{{Hu} \& {Dodelson}(2002)}]{HuDodelson2002}
{Hu} W., {Dodelson} S., 2002, \araa, 40, 171

\bibitem[{{Iannuzzi} \& {Dolag}(2011)}]{Iannuzzi2011}
{Iannuzzi} F., {Dolag} K., 2011, \mnras, 417, 2846

\bibitem[{{Kaiser}(1987)}]{Kaiser1987}
{Kaiser} N., 1987, \mnras, 227, 1

\bibitem[{{Kitaura} {et~al}\mbox{.}(2012){Kitaura}, {Gallerani}, \&
  {Ferrara}}]{Kitaura2012}
{Kitaura} F.-S., {Gallerani} S., {Ferrara} A., 2012, \mnras, 420, 61

\bibitem[{{Komatsu} {et~al}\mbox{.}(2011){Komatsu}, {Smith}, {Dunkley},
  {Bennett}, {Gold}, {Hinshaw}, {Jarosik}, {Larson}, {Nolta}, {Page},
  {Spergel}, {Halpern}, {Hill}, {Kogut}, {Limon}, {Meyer}, {Odegard}, {Tucker},
  {Weiland}, {Wollack}, \& {Wright}}]{Komatsu2011}
{Komatsu} E. {et~al.}, 2011, \apjs, 192, 18

\bibitem[{{Kuhlen} {et~al}\mbox{.}(2012){Kuhlen}, {Vogelsberger}, \&
  {Angulo}}]{Kuhlen2012}
{Kuhlen} M., {Vogelsberger} M., {Angulo} R., 2012, Physics of the Dark
  Universe, 1, 50

\bibitem[{{Ma} \& {Bertschinger}(1995)}]{MaBertschinger1995}
{Ma} C.-P., {Bertschinger} E., 1995, \apj, 455, 7

\bibitem[{{Mao} \& {Wu}(2008)}]{Mao2008}
{Mao} X., {Wu} X., 2008, \apjl, 673, L107

\bibitem[{{Naoz} {et~al}\mbox{.}(2011){Naoz}, {Yoshida}, \&
  {Barkana}}]{Naoz2011}
{Naoz} S., {Yoshida} N., {Barkana} R., 2011, \mnras, 416, 232

\bibitem[{{O'Leary} \& {McQuinn}(2012)}]{Oleary2012}
{O'Leary} R.~M., {McQuinn} M., 2012, \apj, 760, 4

\bibitem[{{O'Shea} {et~al}\mbox{.}(2004){O'Shea}, {Bryan}, {Bordner}, {Norman},
  {Abel}, {Harkness}, \& {Kritsuk}}]{OShea:2004}
{O'Shea} B.~W., {Bryan} G., {Bordner} J., {Norman} M.~L., {Abel} T., {Harkness}
  R., {Kritsuk} A., 2004, arXiv:astro-ph/0403044

\bibitem[{{Rhook} {et~al}\mbox{.}(2009){Rhook}, {Geil}, \&
  {Wyithe}}]{Rhook2009}
{Rhook} K.~J., {Geil} P.~M., {Wyithe} J.~S.~B., 2009, \mnras, 392, 1388

\bibitem[{{S{\'a}nchez} {et~al}\mbox{.}(2008){S{\'a}nchez}, {Baugh}, \&
  {Angulo}}]{Sanchez2008}
{S{\'a}nchez} A.~G., {Baugh} C.~M., {Angulo} R.~E., 2008, \mnras, 390, 1470

\bibitem[{{Scoccimarro}(1998)}]{Scoccimarro1998}
{Scoccimarro} R., 1998, \mnras, 299, 1097

\bibitem[{{Somogyi} \& {Smith}(2010)}]{Somogyi2010}
{Somogyi} G., {Smith} R.~E., 2010, \prd, 81, 023524

\bibitem[{{Springel} {et~al}\mbox{.}(2005){Springel}, {White}, {Jenkins},
  {Frenk}, {Yoshida}, {Gao}, {Navarro}, {Thacker}, {Croton}, {Helly},
  {Peacock}, {Cole}, {Thomas}, {Couchman}, {Evrard}, {Colberg}, \&
  {Pearce}}]{Springel2005a}
{Springel} V. {et~al.}, 2005, \nat, 435, 629

\bibitem[{{Springel} {et~al}\mbox{.}(2001){Springel}, {Yoshida}, \&
  {White}}]{Springel2001a}
{Springel} V., {Yoshida} N., {White} S.~D.~M., 2001, New Astronomy, 6, 79

\bibitem[{{Teyssier}(2002)}]{Teyssier:2002}
{Teyssier} R., 2002, \aap, 385, 337

\bibitem[{{van Daalen} {et~al}\mbox{.}(2011){van Daalen}, {Schaye}, {Booth}, \&
  {Dalla Vecchia}}]{vanDaalen2011}
{van Daalen} M.~P., {Schaye} J., {Booth} C.~M., {Dalla Vecchia} C., 2011,
  \mnras, 415, 3649

\bibitem[{{Wang} \& {White}(2007)}]{Wang2007}
{Wang} J., {White} S.~D.~M., 2007, \mnras, 380, 93

\bibitem[{{White} {et~al}\mbox{.}(2010){White}, {Pope}, {Carlson}, {Heitmann},
  {Habib}, {Fasel}, {Daniel}, \& {Lukic}}]{White2010}
{White} M., {Pope} A., {Carlson} J., {Heitmann} K., {Habib} S., {Fasel} P.,
  {Daniel} D., {Lukic} Z., 2010, \apj, 713, 383

\bibitem[{{Yoshida} {et~al}\mbox{.}(2003){Yoshida}, {Sugiyama}, \&
  {Hernquist}}]{Yoshida2003}
{Yoshida} N., {Sugiyama} N., {Hernquist} L., 2003, \mnras, 344, 481

\bibitem[{{Zel'Dovich}(1970)}]{Zeldovich1970}
{Zel'Dovich} Y.~B., 1970, \aap, 5, 84

\bibitem[{{Zhan} {et~al}\mbox{.}(2008){Zhan}, {Wang}, {Pinto}, \&
  {Tyson}}]{Zhan2008}
{Zhan} H., {Wang} L., {Pinto} P., {Tyson} J.~A., 2008, \apjl, 675, L1

\end{thebibliography}

\label{lastpage} \end{document}